\documentclass[a4paper]{jpconf}
\usepackage{graphicx}
\usepackage{subfigure}
\usepackage{amssymb}
\usepackage[figuresright]{rotating}
\usepackage[rightcaption]{sidecap}
\begin{document}
\title{Azimuthal angular correlations between heavy-flavour decay particles and charged hadrons in pp collisions in ALICE}

\author{Deepa Thomas for the ALICE Collaboration}

\address{ERC-Research Group QGP-ALICE, Utrecht University, Princetonplein 5, 3584 CC Utrecht, The Netherlands}

\ead{deepa.thomas@cern.ch}

\begin{abstract}
Heavy quarks produced in pp and heavy ion collisions are studied using heavy-flavour decay electrons and heavy-flavour mesons. Detailed understanding of the production processes and fragmentation of heavy quarks can be obtained by studying the azimuthal angular correlation of heavy-flavour hadrons. The azimuthal angular correlations of heavy-flavour mesons and charged hadrons can be used to disentangle charm and beauty-hadrons in pp collisions. In this contribution the fraction of electrons from beauty-hadron decays in the heavy-flavour decay electron yield  is shown as well as the beauty production cross section in pp collisions at $\sqrt\textrm{s}$ = 2.76 TeV. The measurements are compared to the predictions from next-to leading order perturbative QCD calculations.  We also show the results from correlation analysis of charged $\mathrm{D}^{*}$ mesons and hadrons performed using pp collision data at $\sqrt{\textrm{s}}$ = 7 TeV. 
\end{abstract}

\section{Introduction}
In heavy-ion collisions heavy quarks are produced by the initial hard scattering processes and probe the high energy density medium produced in such collisions. Heavy flavour mesons can be studied via their hadronic or semi-leptonic decay channels. Detailed understanding on the interaction of heavy quarks in the medium can be achieved by studying the azimuthal angular correlations of heavy-flavour mesons. Correlation studies can be used to obtain information on in-medium partonic energy loss and modification of fragmentation function. In pp collisions, heavy-flavour correlations can be used to disentangle charm and beauty and to test the prediction of perturbative Quantum ChromoDynamic (pQCD) calculations. The measurement in pp is also a necessary baseline for Pb-Pb studies. 

The azimuthal angular correlation between heavy-flavour decay electrons and charged hadrons and between $\mathrm{D}^{*}$ mesons and charged hadrons are presented here. The shape of the azimuthal angular correlations of heavy-flavour decay electrons and charged hadrons are used to determine the relative beauty contribution. Due to different decay kinematics of $\mathrm B$ and $\mathrm D$ mesons, the width of the near side correlation distribution is larger for $\mathrm B$ mesons compared to $\mathrm D$ mesons. Using the heavy-flavour electron cross section and relative beauty contribution to the heavy-flavour electron yield, we can derive the charm and beauty production cross sections separately.

\section{Azimuthal angular correlation between heavy-flavour decay electrons and charged hadrons}
\subsection{Data sample and trigger selection}
The analysis is performed using pp collisions at 2.76 TeV centre-of-mass energy collected in March 2011 with the ALICE experiment ~\cite{aliceDet}. For the analysis, the detectors used are the Inner Tracking System (ITS) ($|\eta|<0.9$, $0<\phi<360^{\circ}$), the Time Projection Chamber (TPC) ($|\eta|<0.9$, $0<\phi<360^{\circ}$) and the Electromagnetic Calorimeter (EMCal) ($|\eta|<0.7$, $80<\phi<180^{\circ}$). The events which pass EMCal L0 trigger conditions are used. The L0 trigger is a 2$\times$2 tower patch with a trigger cluster energy threshold of 3 GeV.

\subsection{Data Analysis}
Electrons are identified using information from the TPC and EMCal detectors. Particle identification in the TPC is based on the measurement of the specific ionization energy loss in the detector gas. In the EMCal, electron candidates are required to have $E/p$ between 0.8 and 1.2. The non-heavy-flavour electrons (Non-HFE) are mainly produced by internal and in-material conversions and are identified using the invariant mass method, where the partner electron is selected with loose electron identification criteria. Electron pairs which have an invariant mass of less than 50 MeV/$c^{2}$ are tagged as Non-HFE. Above an electron $p_{\rm T}$ of 2 GeV/$c$, the non-HFE finding efficiency is $\approx 50\%$, as estimated from Monte Carlo (MC) simulations. The remaining Non-HFE contamination in the heavy flavour electron (HFE) sample is corrected for using this efficiency.

\begin{SCfigure}[1.2][h]
\centering
\includegraphics[height=6cm, width=6.5cm]{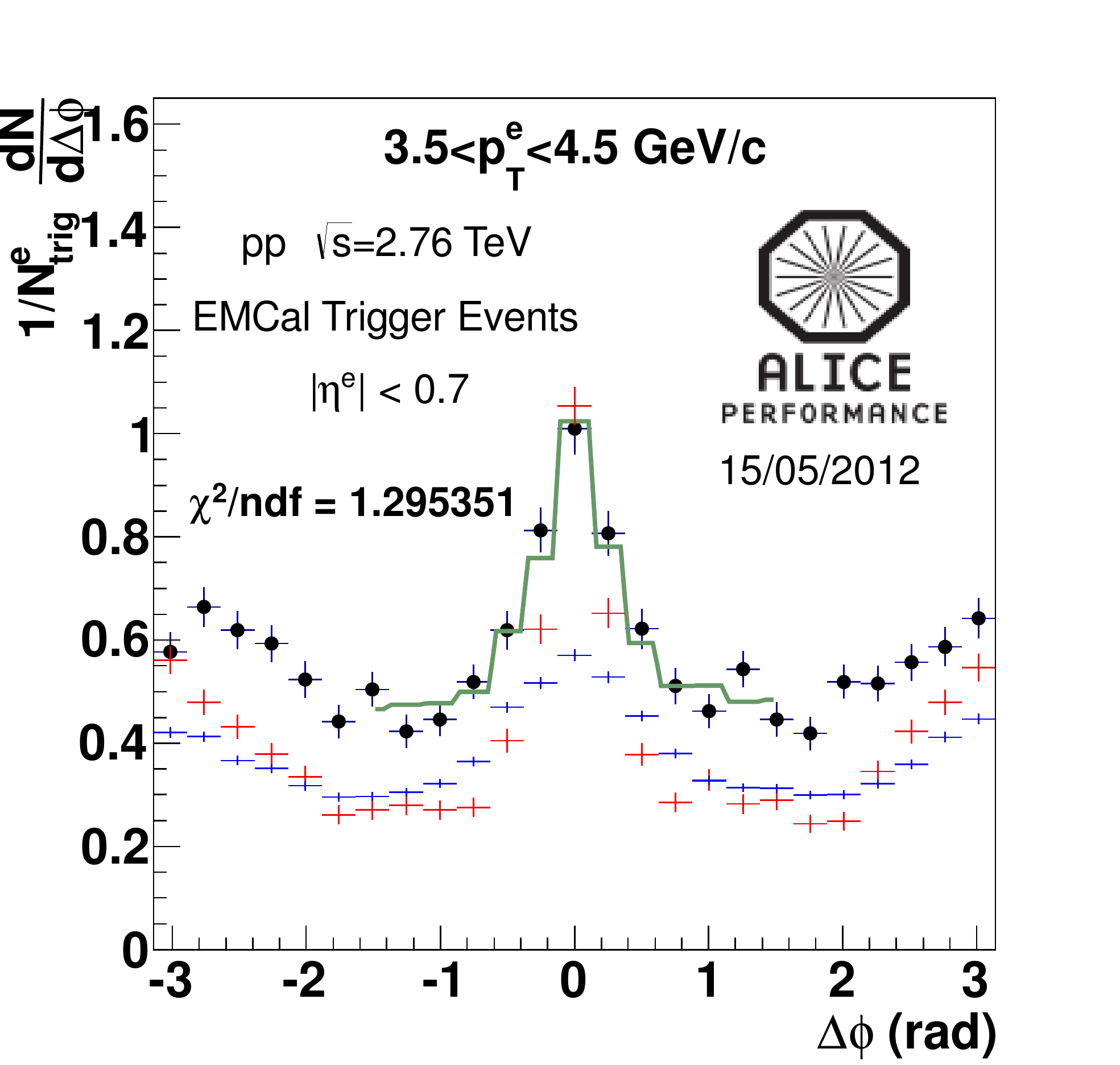}
\label{fig:deltaphifit}
\hspace{0.3cm}
\caption{Azimuthal angular correlation between heavy-flavour decay electrons and charged hadrons in pp collisions at 2.76 TeV. The MC distribution for electrons from charm decay is shown in red, while the MC distribution for electrons from beauty decay is shown in blue and the full green curve is the fit to the data points.}
\end{SCfigure}

\begin{figure}[h]
\centering
  \subfigure{
        \includegraphics[height=6cm,width=7.5cm]{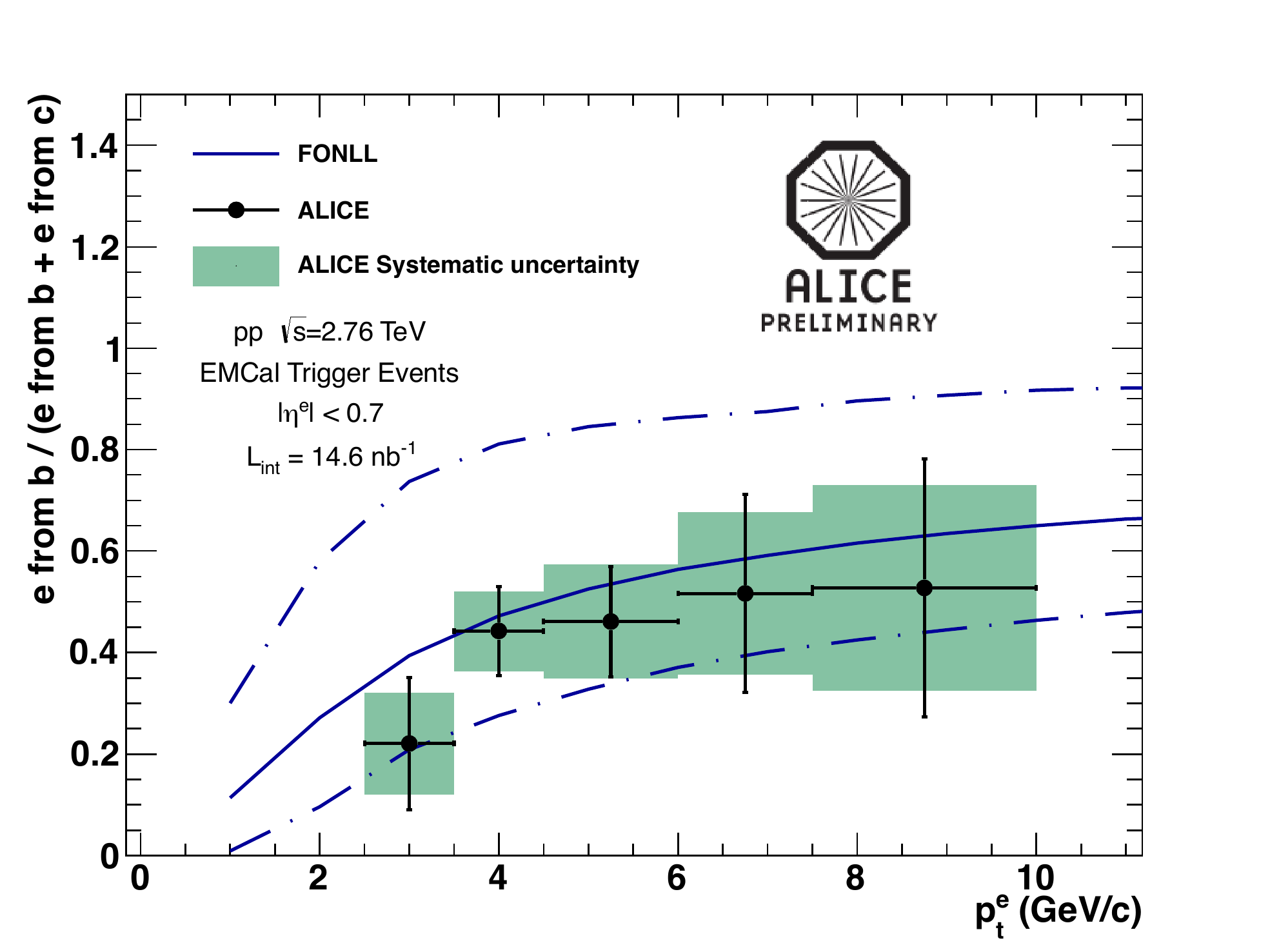}
  }
\subfigure{
          \includegraphics[height=6cm,width=7.5cm]{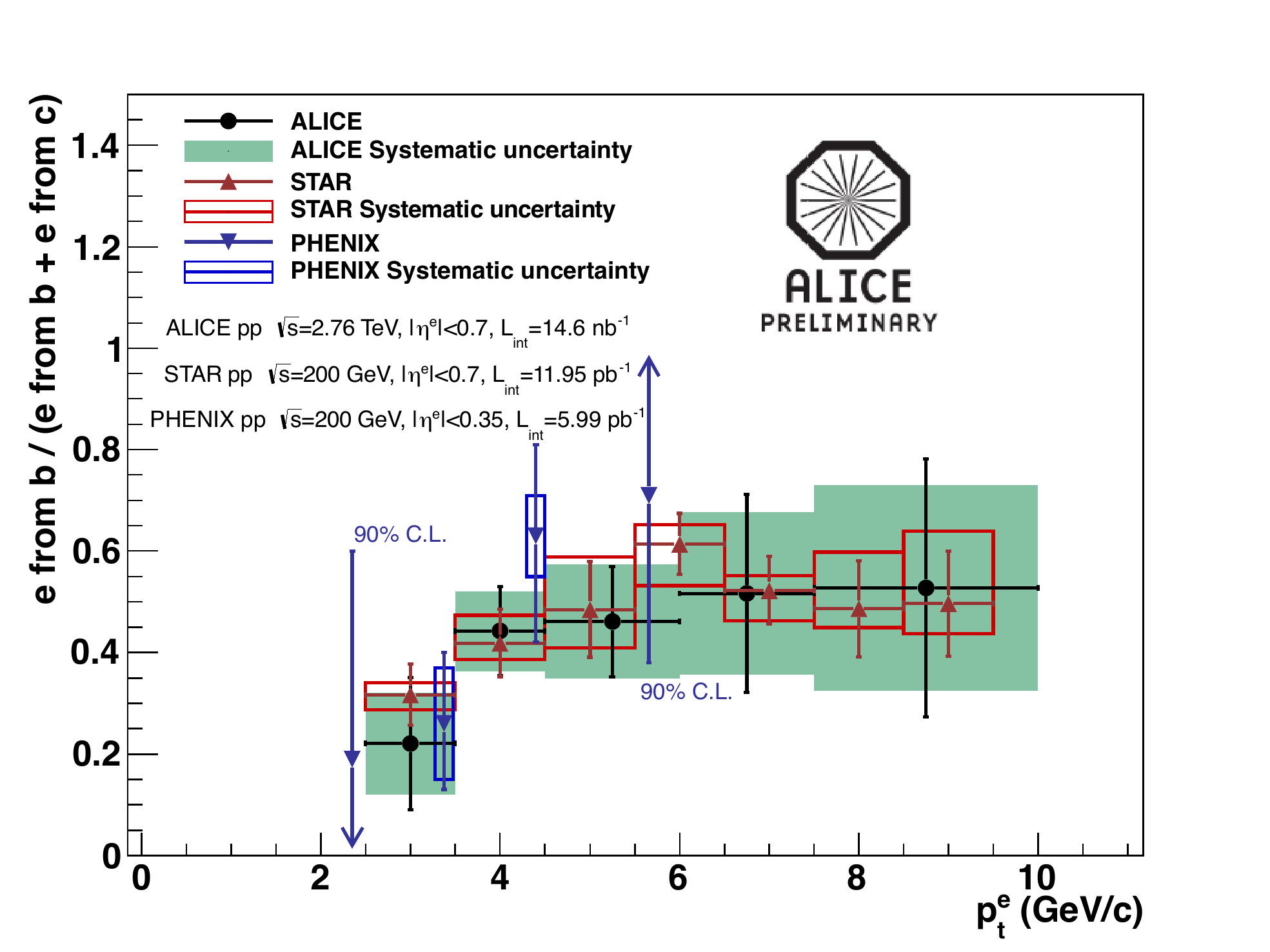}
}
\caption[]{Relative beauty contribution to the heavy-flavour decay electron yield in pp collisions at $\sqrt{\textrm{s}} =$ 2.76 TeV compared with FONLL calculations ~\cite{FONLLcurve} (left) and previous RHIC measurements ~\cite{BottomStar,eh-Bcs-Phenix} (right). }
\label{fig:bratio}
\end{figure}

The azimuthal angular correlations between heavy-flavour decay electron and charged hadrons is constructed using high quality tracks. In order to determine the fraction of electrons from beauty decays $r_{\mathrm B}$, the measured correlation distribution is fit with the function

\begin{equation}
\Delta\phi_{\mathrm {e-h}}^{\mathrm {HF}} = const + r_{\mathrm {B}} \Delta\phi_{\mathrm {e-h}}^{\mathrm {B}} + (1-r_{\mathrm {B}}) \Delta\phi_{\mathrm{e-h}}^{\mathrm D},
    \end{equation}

    where  $r_{\mathrm B}$ = $\frac {\mathrm {e_{B}}} {\mathrm{e_{B}} + \mathrm{e_{D}}}$ is the ratio of electron yield from B-meson decays to that of the heavy-flavour electron yield, $\Delta\phi_{\mathrm{e-h}}^{\mathrm{D}}$ ($\Delta\phi_{\mathrm{e-h}}^{\mathrm{B}}$ ) is the azimuthal angular correlation between electrons from D (B) meson decay and charged hadrons from MC simulations (PYTHIA 6.4 with Perugia-0 tune ~\cite{Pythiatune}) including detector response. $Const$ term describes the uncorrelated background.

    The fitting range used is $-1.5 < \Delta\phi < 1.5$ rad. The correlation distribution and the fit are shown in Figure \ref{fig:deltaphifit}. The beauty fraction extracted from the fit as a function of $p_{\rm T}$ is shown in Figure \ref{fig:bratio}. The $r_{\mathrm{B}}$ increases with $p_{\rm T}$ and reaches $\approx 0.5$ ($\frac{\mathrm{e_{B}}}{\mathrm{e_{D}}} \approx 1$) at around 5 GeV/$c$. This measurement is consistent, within uncertainities, with perturbative Quantum ChromoDynamics Fixed Order plus Next-to-Leading Logarithms (pQCD FONLL) calculations ~\cite{FONLLcurve} and  with results from RHIC in pp collisions at $\sqrt{\textrm{s}} =$ 200 GeV ~\cite{BottomStar,eh-Bcs-Phenix}.

\section{Beauty and charm decay electron cross section}
The beauty and charm decay electron cross section are computed using the heavy-flavour decay electron cross section measured in ALICE and the relative beauty contribution. The heavy-flavour decay electron cross section is measured using the same data sample as the correlation analysis and the procedure described in ~\cite{7TeVHFE}. Electrons are measured using the TPC and EMCal detectors. The cross section is measured after applying various corrections like tracking efficiency and unfolding, particle identification efficiency and purity and EMCal trigger efficiency. The Non-HFE background is identified and subtracted using a cocktail method where the measured $\eta$ and $\pi^{0}$ cross section are used as an input and all the known sources of background electrons are taken into account ~\cite{7TeVHFE}. 

\begin{figure}[h]
\centering
  \subfigure{
    \includegraphics[scale=0.25]{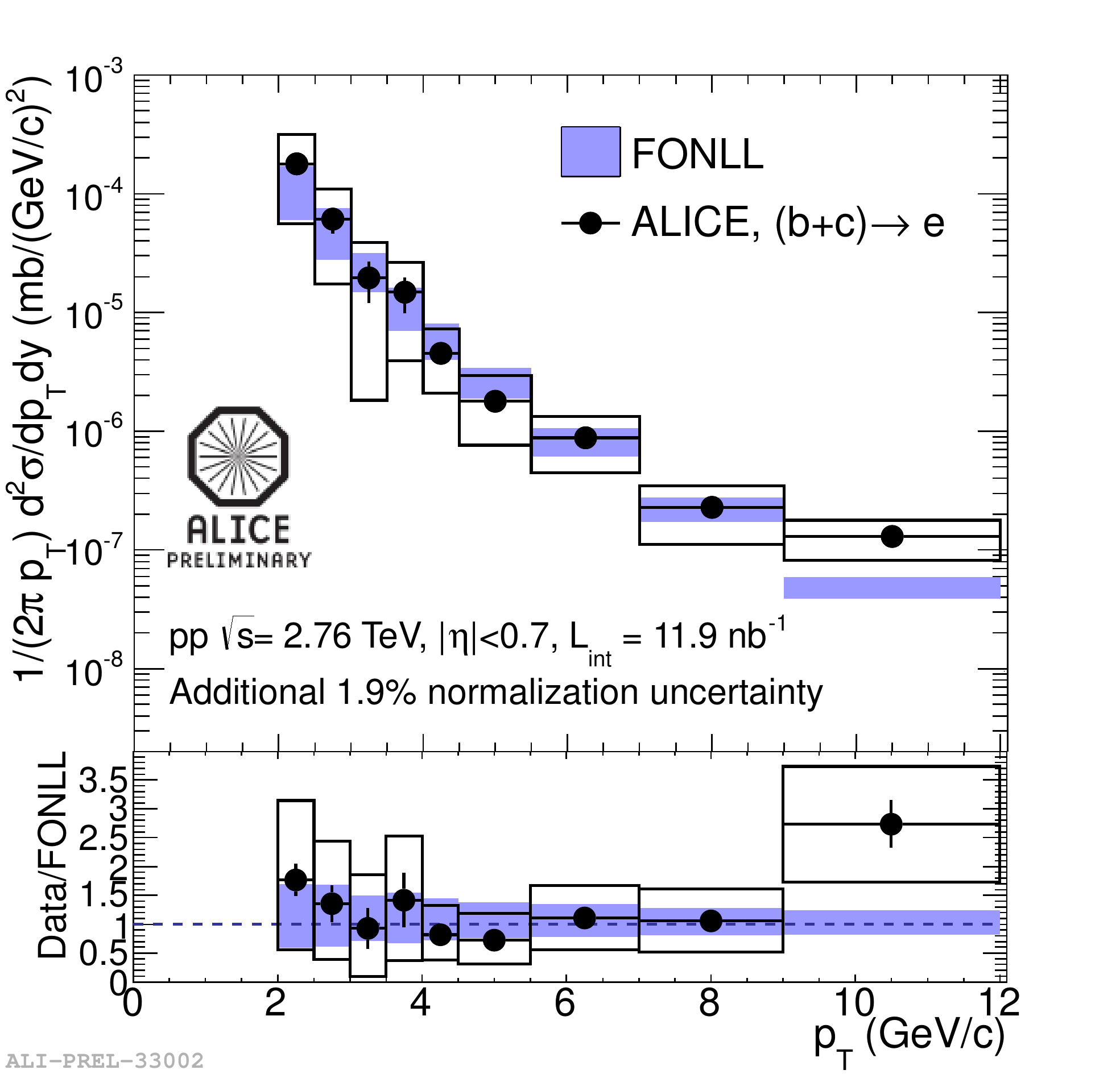}
      \label{fig:hfe crosssection}                                                                      
  }
  \subfigure{
    \includegraphics[scale=0.25]{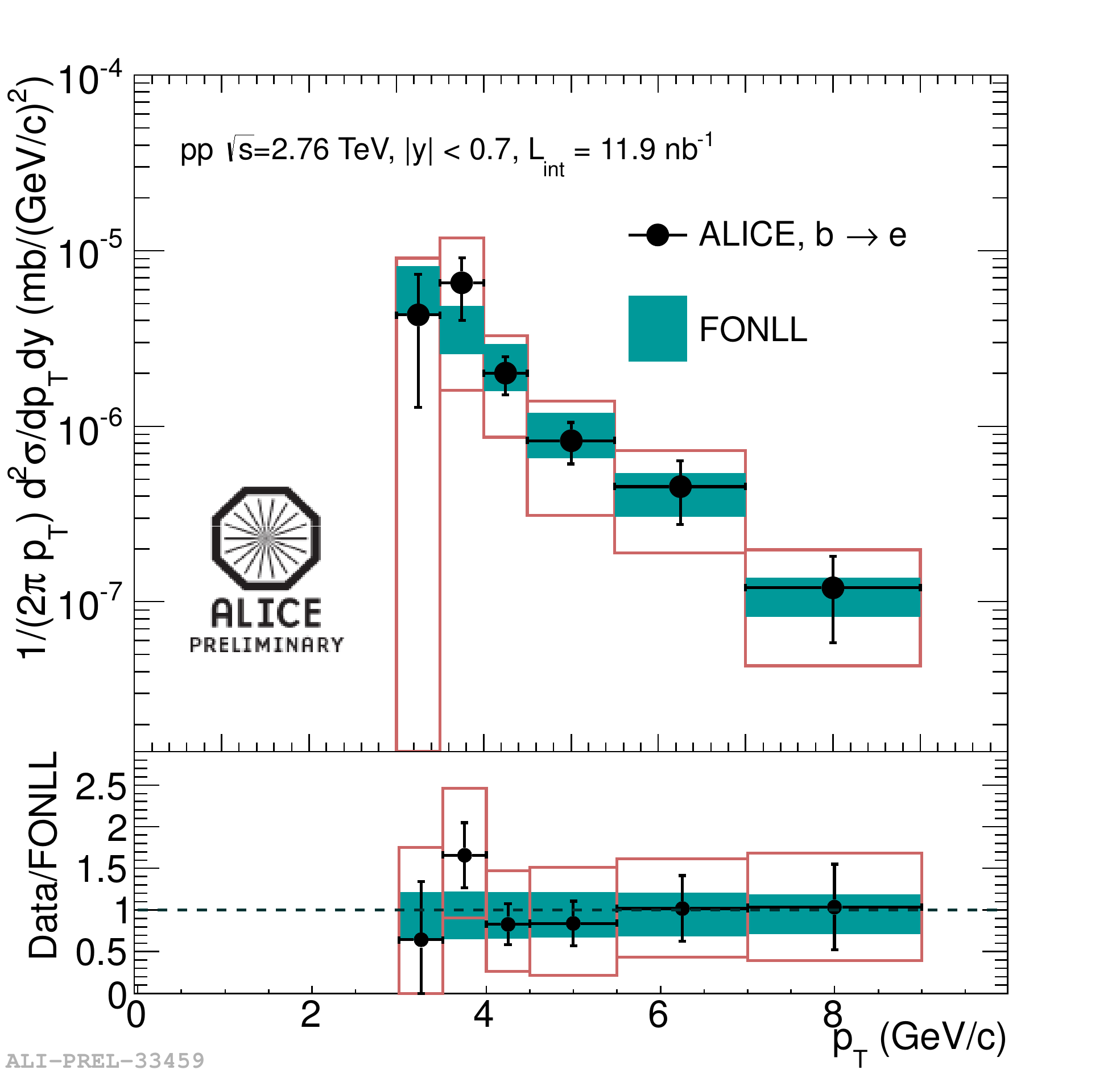}
    \label{fig:b cross section}
  }
  \subfigure{
    \includegraphics[scale=0.25]{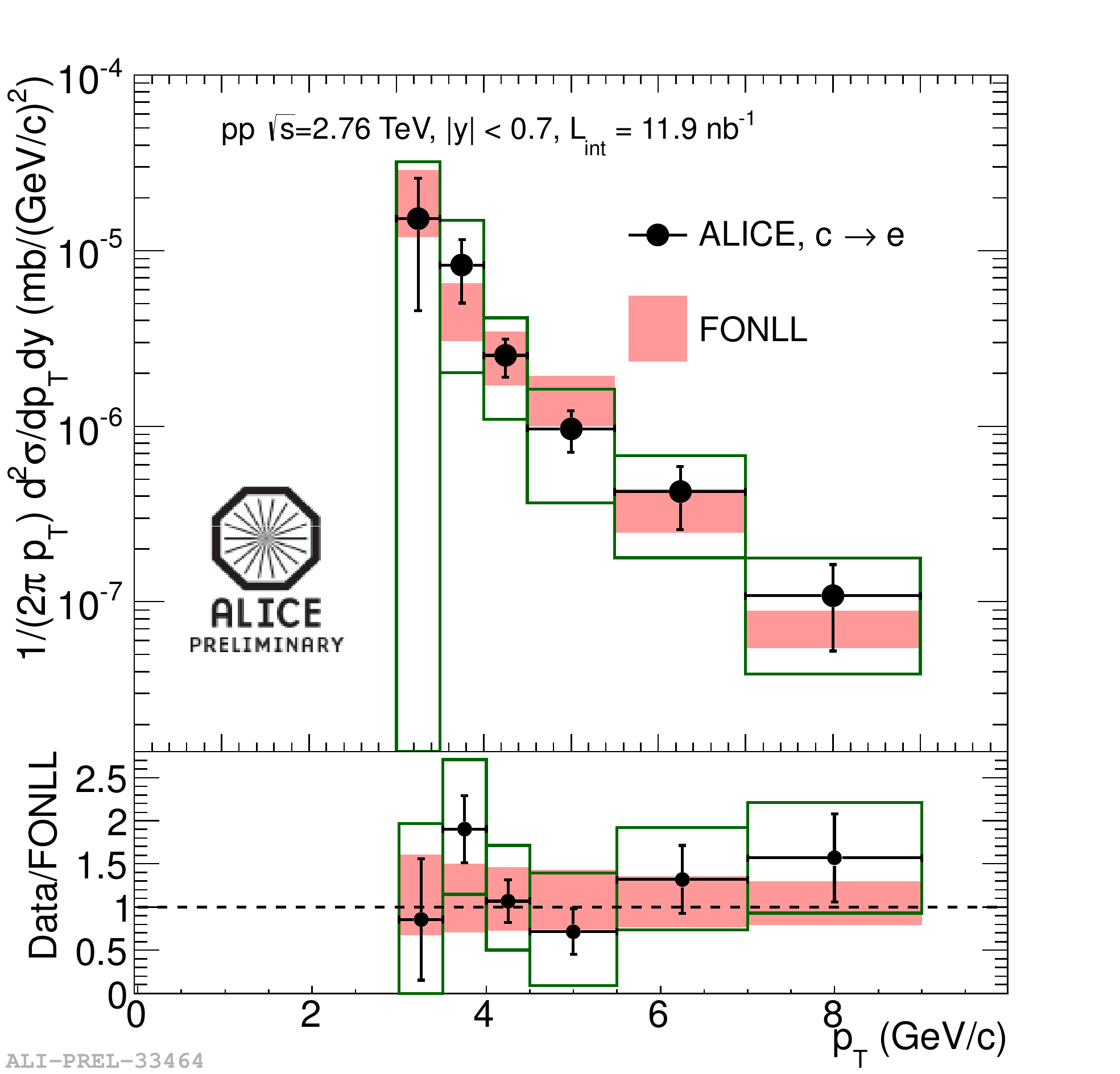}
    \label{fig:b cross section}
  }

\caption[]{Heavy flavour decay electron (left), beauty (middle) and charm (right) decay electron cross section in pp collisions at $\sqrt{\textrm{s}} =$ 2.76 TeV compared to FONLL calculations.}
\label{fig:Cross-section}
\end{figure}

The beauty and charm decay electron cross section is computed as

\begin{equation}
\left(\frac{\textrm{d}\sigma}{\textrm{d}p_{\rm T}}\right)_{\mathrm{b}\rightarrow \mathrm{e}} = r_{\mathrm{B}} \times \left(\frac{\textrm{d}\sigma}{\textrm{d}p_{\rm T}}\right)_{\mathrm{b+c}\rightarrow \mathrm{e}}
\end{equation}

\begin{equation}
\left(\frac{\textrm{d}\sigma}{\textrm{d}p_{\rm T}}\right)_{\mathrm{c}\rightarrow \mathrm{e}} = \left(\frac{\textrm{d}\sigma}{\textrm{d}p_{\rm T}}\right)_{\mathrm{b+c}\rightarrow \mathrm{e}} - \left(\frac{\textrm{d}\sigma}{\textrm{d}p_{\rm T}}\right)_{\mathrm{b}\rightarrow \mathrm{e}}
\end{equation}

since  $r_{\mathrm{B}}$ is not measured in the full $p_{\rm T}$ range of the HFE cross section, the beauty and charm decay electron cross section is measured from 3 to 9 $\textrm{GeV}/c$. The measured heavy-flavour electron and beauty and charm decay electron cross section is shown in Figure \ref{fig:Cross-section} together with FONLL pQCD calculations ~\cite{FONLLcurve}. The cross section are consistent with FONLL calculations. 

\section{Azimuthal angular correlation between $\mathrm{D}^{*}$ mesons and charged hadrons}

\begin{figure}[h]
\centering
\subfigure{
  \includegraphics[scale=0.27]{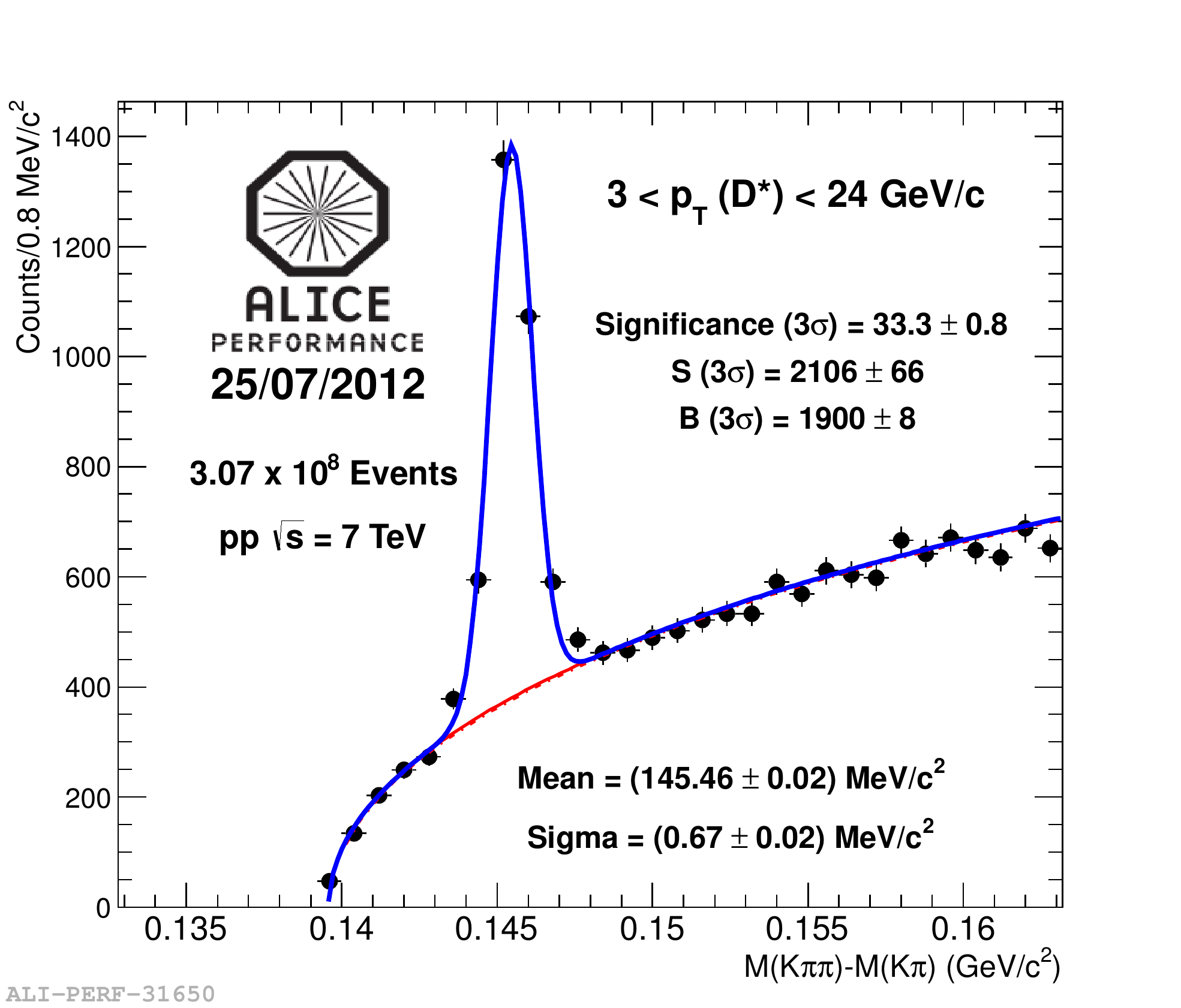}
  \label{fig:D*reco}
}
\subfigure{
  \includegraphics[scale=0.24]{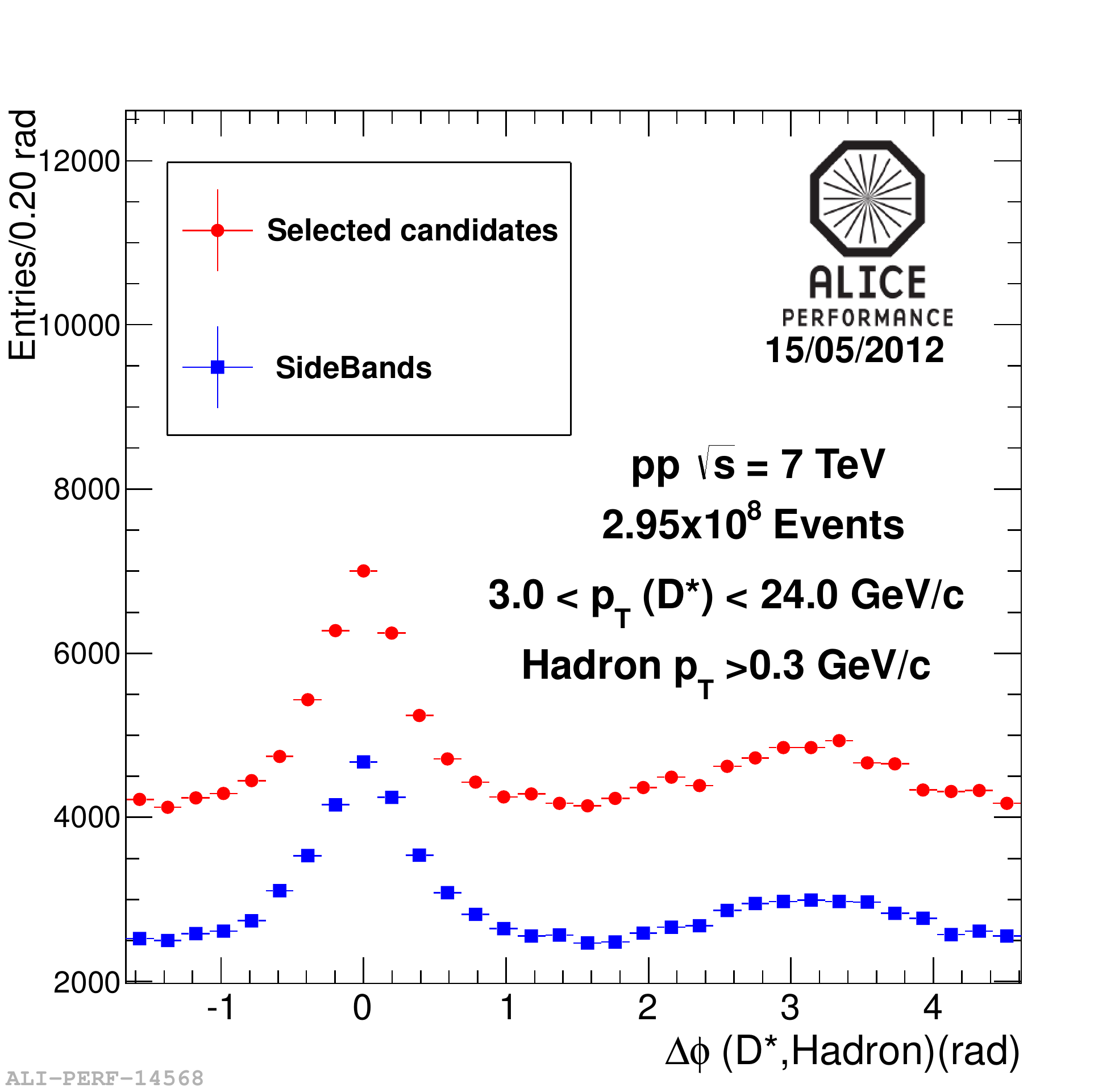}
  \label{fig:D*Correl}                                                           
}
\caption[]{(Left) Invariant mass distribution of $\mathrm{D}^{*}$ mesons. (Right) Azimuthal angular correlation between $\mathrm{D}^{*}$ and charged hadrons in pp collisions at $\sqrt{\textrm s} =$ 7 TeV.}
\label{fig:D*}
\end{figure}

The correlation between charged $\mathrm{D}^{*}$ and charged hadrons is performed in pp collisions at $\sqrt{\textrm s}=$ 7 TeV collected in 2010. $\mathrm{D}^{*}$ mesons are reconstructed via the decay channel $\mathrm{D}^{*}\rightarrow\mathrm{D}^{0}(K\pi)\pi$ using the invariant mass method, see Figure \ref{fig:D*} (left). The $\mathrm{D}^{*}$ background is estimated using $\mathrm{D}^{0}$ background candidates selected in the invariant mass side bands $4\sigma<|\mathrm{M}(K\pi)-\mathrm{M}(\mathrm{D}^{0})|<10\sigma$ and combining this fake $\mathrm{D}^{0}$ with pion tracks. The $\mathrm{D}^{*}$ mesons are correlated in azimuth with hadrons which pass quality track selection criteria. The reconstructed $\mathrm{D}^{*}$ mesons and the correlation distribution of $\mathrm{D}^{*}$ mesons with charged hadrons are shown in Figure \ref{fig:D*} (right). The red distribution gives the azimuthal angular correction for all $\mathrm{D}^{*}$ candidates and the blue distribution corresponds to sideband background candidates. The distribution can be fitted with gaussian distribution to extract the correlation parameters (yield and width).

\section{Results}
The relative beauty contribution to the heavy-flavour decay electron yield was measured in pp collisions at $\sqrt{\textrm s} =$ 2.76 TeV with the ALICE detector and compared with pQCD calculations and RHIC measurements. Using this beauty fraction, the beauty-decay and charm-decay electron cross sections are derived and they are found to be consistent with FONLL pQCD calculations.
\section{References}

\end{document}